\documentstyle[epsfig,aps,prl,floats,twocolumn]{revtex}
\begin{document}

\title{Superconductor/Insulator Transition in the Striped Phase}
\author{Yu. A. Dimashko and C. Morais Smith}
\address{I Institut f{\"u}r Theoretische Physik, Universit{\"a}t 
Hamburg, D-20355 Hamburg, Germany \\ 
\rm{(\today)}\bigskip\\
\parbox{14.4cm}{\rm
We study the transversal dynamics of a charged stripe (quantum string) at 
 $T = 0$ and show that its kink excitations play the role of current carriers. 
If the hopping amplitude $t$ is much smaller than the string tension $J$, the 
string is pinned and its ground state (GS) is insulating. At $t \gg J$,
the string is depinned and the GS is a kink-condensate. By mapping the system
onto a Josephson junction chain, we show that this state
is superconducting. At $(t/J)_c = 2 / \pi^2 $ the kink/antikink pairs decouple
and a Kosterlitz-Thouless like insulator-superconductor transition occurs. 
 }}

\maketitle
\narrowtext

The existence of a striped phase in doped 2D antiferromagnets (AF) has been recently
a subject of intense experimental \cite{Tran,Yama,Bors}, numerical 
\cite{Zoto,Giam,Scal}, and  theoretical \cite{theo} investigations. 
Experimentally, elastic \cite{Tran} and inelastic \cite{Yama} neutron 
diffraction measurements in nickelates and cuprates have revealed the presence
of charge and spin-order. Besides, muon spin resonance and nuclear quadrupole 
resonance results \cite{Bors} have been also successfully interpreted within 
the picture of charged domain walls separating antiferromagnetic domains. 
Numerically, several studies reproduce this finding: in the framework of the 
one- or three-bands Hubbard model, or the simpler $t-J$ model, exact
diagonalization in small clusters \cite{Zoto}, quantum Monte Carlo 
\cite{Giam}, and density matrix renormalization group \cite{Scal} 
studies show the existence of the stripe configuration. 
Theoretical calculations based on the Hartree-Fock approximation \cite{theo} 
also point in the same direction. 

The stripe configuration in doped AF can be regarded as an extended 
 charge density wave (CDW). 
The conductivity from CDW was considered initially by Fr{\"o}hlich \cite{Froh}, 
who showed the perfect conductivity of the collective (sliding) mode. The physical 
cause of this phenomenon is the blocking of the scattering processes by the Landau criterion
\cite{Land}.  Later, Lee, Rice, and Anderson (LRA) \cite{LRA} analyzed the dynamics of the
sliding mode in the presence of a lattice. They argued that pinning by the lattice and/or by
impurities destroys the perfect conductivity. However, even pinned,  the sliding mode gives
rise to  a large dielectric constant, as was experimentally observed in  quasi 1D-compounds
\cite{expcdw}.

It is important to notice that LRA did not consider quantum fluctuations. This  factor can
compete with pinning even at $T=0$. The question is, whether the quantum fluctuations are able
to depin the stripe (or CDW), and whether this quantum depinning restores the perfect
conductivity of the sliding mode.

In the present Letter, we investigate this question within a phenomenological model \cite{Eske}, 
which can be related to the $t-J$ model \cite{Mora}. We study the 
transversal dynamics of a single stripe, which is treated as 
a quantum string. Then, by performing a canonical transformation in the 
quantum string Hamiltonian, we map the system onto a 1D array of Josephson 
junctions, which is known to exhibit an insulator/superconductor 
transition at $(t/J)_c =2 / \pi^2$. Further, we calculate the ground state (GS) of the
quantum string in two limiting cases, $t \ll J$ and $t \gg J$, and reveal the meaning of the
transition in the ``string'' language. We argue that kink excitations of the string play the 
role of current carriers. At $(t/J)_c$  the {\it insulating} pinned phase, corresponding to
a GS with bound kink/anti-kink (K/AK) excitations turns into a {\it superconducting}
depinned phase, where the K/AK pairs are decoupled and form a kink-condensate.
In doing so, we have connected two important and different classes of 
problems, i.e., the transversal dynamics of stripes in doped AF and a system 
with well known superconducting properties. This suggests that phase coherence
can be established within the stripes, providing us with an important clue
for the understanding of the superconductivity mechanism in high-$T_c$
superconductors. 

Let us consider a single vertical string on a $N\times M$ square lattice  
(see Fig.\ 1a). The linear concentration of holes in the string is assumed to 
be one hole/site. The string is composed of $N$ charged particles elastically 
interacting with the neighbour ones and constrainted to move along $N$ 
horizontal lines. The lattice constant is taken as the unit of length. 

The classical state of the system is described by the  $N$-dimensional vector 
 ${\vec x}= \{x_1,x_2,...,x_N\}$. 
Here, $x_n$ is the $x$-coordinate of the $n$-th particle, $x_n=1,2,...,M$. 
The corresponding quantum state $\mid \vec x >$ is 
defined as an eigenstate of all the coordinate operators  
 $\hat{x}_n$, $n=1,2,...,N$ : $\hat{x}_n \mid \vec{x} > = x_n \mid \vec{x}>$.
The phenomenological Hamiltonian describing this system is 
\begin{equation}
\hat{H}=-t\sum_n \left(\hat{\tau}_n^{+} + \hat{\tau}_n^- \right) + 
\frac{J}{2}\sum_n \left( \hat{x}_{n+1} - \hat{x}_n \right)^2.
\label{Ham1}
\end{equation}

\begin{figure}[t]
\unitlength1cm
\vspace{0.1cm}
\begin{picture}(8,6)
\epsfxsize=5cm
\put(1.0,0.5){\epsfbox{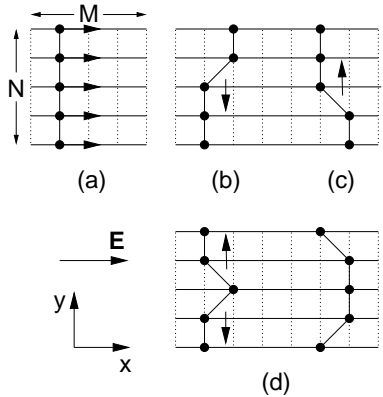}}
\end{picture}
\caption[]{\label{fig1} Localized states of the quantum string and their translation modes
induced by the electrical field ${\bf  E}$: a) flat state; b) kink; c) antikink; 
d) pair kink/antikink .}
\end{figure}

The translation operators $\hat{\tau}_n^{\pm}$ are defined
by their action on the coordinate states, 
$\hat{\tau}_n^{\pm} \mid \vec{x} > = \mid \vec x \pm \vec{e}_n >$, 
where  $(\vec{e}_n)_m = \delta _{nm}$. 
The coefficients $t$ and $J$ denote the hopping amplitude and the string 
tension, respectively. The operators $\hat{\tau}_n^{\pm}$ can be expressed 
through the momentum operators $\hat{p}_n$, which obey the canonical relation 
 $[\hat{x}_n,\hat{p}_m ] = i \delta_{nm}$.
We then find $\hat{\tau}_n^{\pm}=\exp (\pm i \hat{p}_n )$ and 
the Hamiltonian (\ref{Ham1}) becomes 
\begin{equation}
\hat{H}=-2t\sum_n \cos \hat{p}_n + \frac{J}{2}
\sum_n \left(\hat{x}_{n+1}-\hat{x}_n \right)^2.
\label{Ham2}
\end{equation}

Hereafter, we classify the state of the quantum string by the value of the topological charge
$\hat{Q} = \sum_n ( \hat{x}_{n+1}- \hat{x}_n).$ In the case of open boundary conditions
(BC), the topological charge is an arbitrary integer, $Q=0,\pm 1,\pm 2,...$ The states with
positive and negative charges are called kinks (K) and antikinks (AK), respectively (see Figs.\
1b and 1c). Here, we consider periodic BC, $\hat{x}_{N+1}= \hat{x}_1$. Hence, 
the total topological charge of the string is zero (Figs.\ 1a and 1d).

Since we are interested in the conducting properties of the system, we have
to determine the current operator $\hat{\j}_n = e \dot{\hat{x}}_n$,
where $e$ is the charge of the particle and the dot denotes the time derivative. 
Using the equation of motion $\dot{\hat{x}}_n = i[\hat{H}, \hat{x}_n]$, 
we obtain $\hat{\j}_n = 2 e t \sin \hat{p}_n$.

At this point, it is convenient to perform a dual transformation to new
variables refering to the segments of the string, i.e., to a pair of
neighbour holes, 
\begin{equation}
\hat{x}_n - \hat{x}_{n-1} = \hat{\pi}_n, \qquad 
\hat{p}_n = \hat{\varphi}_{n+1} - \hat{\varphi }_n.
\label{dt}
\end{equation}
The new local variables also obey the canonical relation
 $[\hat{\varphi}_n, \hat{\pi}_m ] = i \delta_{nm}$. Furthermore,  we take the limit $M \to
\infty$ in order to deal with all operators  in the $\varphi $-representation, $ \hat{\varphi }_n
\Rightarrow \varphi _n,\, \hat{\pi}_n \Rightarrow -i\partial /\partial \varphi _n$.
The continuous variable $\varphi_n$ is restricted to the interval
 $0 \leq \varphi_n < 2 \pi$. Finally, the Hamiltonian and the current 
operator acquire the form
\begin{eqnarray}\nonumber
\hat{H}=-2t\sum_n\cos (\varphi_{n+1}-\varphi_n) &-& \frac{J}{2}
\sum_n(\partial /\partial \varphi_n)^2, \\
\hat{\j}_n=2et\sin (\varphi_{n+1} &-& \varphi_n),
\label{seg}
\end{eqnarray}
which is known from the theory of superconducting chains.
Eqs.\ (\ref{seg}) describe a Josephson junction chain, with the
Coulomb interaction taken into account. The solution of this problem
at $T = 0$ has been found by Bradley and Doniach \cite{Brad}. Depending
on the ratio $t/J$, the  chain is either insulating (small $t/J$) or
superconducting (large $t/J$). The results arise from the standard mapping
of the 1D quantum problem onto the 2D classical one. In this way, one obtains
the XY model with Euclidean action 
\begin{equation}
{\cal S}_E = \sqrt{\frac{2t}{J}} \sum_{<\vec{r}, \vec{r}'>} \cos 
\left( \varphi_{\vec{r}} - \varphi_{\vec{r}'} \right),
\label{Ecaction}
\end{equation}
where the vectors $\vec{r} = (n, \tau)$ form a rectangular lattice 
in space and imaginary time.

At $t/J = 2/ \pi^2$ the Josephson chain undergoes a Kosterlitz-Thouless (KT) 
transition. For
small $t/J$  values, the two-points correlator 
 $<\exp  i ( \varphi_{\vec r} - \varphi_{\vec r'})  >$ decays exponentially.
Then, the frequency dependent conductivity exhibits a resonance, 
Re $\sigma(\omega) \propto \delta (\omega - J)$. 
Since there is no conductivity at $\omega = 0$, this is an insulating state
with a gap $\Delta = J$. In the opposite case, when $t/J$ is large, the same
correlator decays algebraically. Then, the conductivity is
singular at $\omega = 0$, Re $\sigma (\omega) = 2 \pi e^2 t \delta
(\omega)$, and the system is superconducting. 

These results are also valid for the quantum string on the lattice, since
both systems are described by exactly the same Hamiltonian and 
current operator, see Eqs.\ (\ref{seg}). Now, it remains to reveal the 
physical signification of these results for the striped phase. 
Mainly, it is important to understand the origin of the current and the 
meaning of the KT transition in the ``string language''. In order to achieve 
this aim, it is instructive to analyse the problem in two limiting cases, 
which allow for an explicit solution: $t\ll J$ and $t\gg J$.

Let us start considering the limit of weak fluctuations, with $t\ll J$.
In the absence of hopping, $t=0$, the excitation spectrum of the string is
discrete, $E_\nu = \nu J$, with $\nu = 0, 1, 2...$ 
The ground level  $E_0 = 0$ corresponds to a flat state of the string,
i.e., to the kink-vacuum $\mid 0 >$  (see Fig.\ 1a).
The first elementary excitation corresponds
to the creation of a pair K/AK (see Fig.\ 1d), i.e., to the state 
$\hat{A}^{+}_{n}\hat{A}^{-}_{m}\mid 0 >$ . Here, 
$\hat{A}^{\pm}_{n}= \exp (\pm i \hat{\varphi}_{n})$ are local creation operators of 
K$(+)$ and AK$(-)$. This state has energy $E_1 = J$. All the levels are degenerated, and
therefore, accounting for small but nonzero hopping $t$ can split them.

The ground level $E_0$ splits only in the $N$-th order of the perturbation theory.
Indeed, any flat state of the string can be transformed into another flat one 
only after $N$ elementary translations (see Fig.\ 1a). Hence, the ground level $E_0$ acquires a
width  $\propto J(t/J)^N$, which tends to zero in the thermodynamical limit 
 $N\rightarrow \infty $.  In this  limit the ground level is $not$ split. 

The first excited level $E_1$ splits already in the first order of the perturbation theory. 
Resolving the secular equation, one
finds the energy $E_1(k,q)=J-2t(\cos k+\cos q)$
and the corresponding  states 

\begin{equation}
\mid k,q > =
\frac {1}{\sqrt{2}}(\hat{A}^{+}_{k}\hat{A}^{-}_{q}-\hat{A}^{+}_{q}\hat{A}^{-}_{k})\mid 0 >.  
\label{psi1}
\end{equation}
Here, $A^{\pm} _{k}= (1/\sqrt{N})\sum_n \exp (\pm i\hat{\varphi}_n +ikn)$
are creation operators of K/AK with fixed momentum $k$.

Thus, the level $E_1$ acquires finite band width $\simeq 8t$. Actually, one can
show that each $\nu$-th level splits into a band of width $\simeq 8\nu t$.  The energy band
structure of the quantum string is shown in  Fig. 2. The 
excitation spectrum has a gap $\Delta \simeq J-4t$, which is nothing 
but the minimal energy required to create a K/AK pair. The existence of the 
gap  at $t\ll J$ suggests that the ground state (GS) of the string in this limit is insulating. 

\vspace{1.3cm}
\begin{figure}[ht]
\unitlength1cm
\begin{picture}(7,3.5)
\put(1.5,0.5){\epsfbox{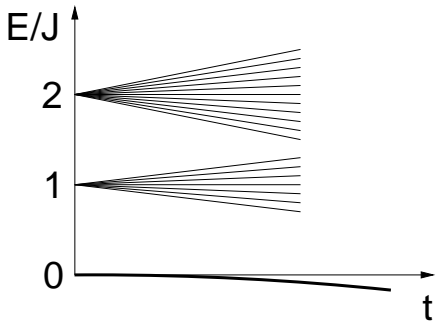}}

\end{picture}

\caption[]{\label{fig4} Energy spectrum of the quantum string at small nonzero hopping, 
$t \ne 0$ and $t / J \ll 1$. The ground level is not split, but it decreases as a quadratic power-law.}
\end{figure}

Nevertheless, it is interesting to consider the conductivity of the {\it excited} states
caused by the presence of the kinks.  In the excited state  (\ref{psi1}) we have one unbound
pair K/AK. They can move in opposite $y$-directions.  This leads to an
effective one-step motion of the string in the $x$-direction, with the consequent appearance of
current (see Fig.\ 1d). Thereby, the kinks play the role  of current carriers. By averaging the
current operator   $\hat{I} = \sum_n \hat{\j}_n$ over the K- and AK-states 
 $\hat{A}^{\pm}_{k} \mid 0 >$, we find $<\hat{I}> = \mp 2et \sin k$, which is similar to
the current $I = 2et \sin k$ for one free hole. However, in contrast to the hole, the 
effective electrical
charge of the kink is determined by its topological charge. This fact can also be seen from a
different analysis.  Let us apply an electrical field $E$ along the $x$ axis (see Fig.\ 1).
Then, a one-step shift of the K (AK) in the $y$-direction, $y \to y + 1$, results in a energy 
change of $eE$ $(-eE)$. Hence, one can ascribe an effective electrical  charge $-e$ $(+e)$ to
the K (AK).

It is worth to note that in the  $t \ll J$ limit the GS has only local K/AK pairs.
A perturbative calculation of the K/AK pair correlation function reveals only short-range
correlations  
$< \hat{A}^{+}_{n}\hat{A}^{-}_{m}> =<\exp  i ( \hat{\varphi}_n - \hat{\varphi}_m )>\propto \exp
(- |n - m|/R)$. The correlation length  $R=2/\ln (J/2t)$ can be treated as an average dimension
of the virtual K/AK pairs. The local K/AK pair cannot carry current because it forms a
bound state with zero topological charge.
Thereby, the locality of the K/AK pairs is reponsible for the insulating character of
the GS at small $t$.

Now, we concentrate on the opposite limit, when the fluctuations are strong
and $t\gg J$. In this case, we can expand the $t$-term in the Hamiltonian 
(\ref{seg}) up to second order, $\cos (\varphi_{n+1} - \varphi_n) \approx 1 - (\varphi_{n+1} -
\varphi_n)^2 /2$, and  diagonalize the quadratic Hamiltonian. Then, we obtain the 
phonon-like spectrum $E_k=-2tN+\omega_k$,  
\begin{equation} 
\omega_k=\sqrt{8tJ}\mid \sin (k/2)\mid,
\label{phonon}
\end{equation}
with a finite band width $\sqrt{8tJ}$ and no gap. Therefore, the GS is
conducting. 
Calculations of the conductivity are straightforward, since in this case, the time dependence of 
the current  $\hat{\j}_n \approx 2et (\hat{\varphi}_{n+1} - \hat{\varphi}_{n})$ 
follows from the standard relation, 
\begin{equation} 
\hat{\varphi}_n (\tau) 
= \sum_k \sqrt{\frac{J}{2N\omega_k}}
[e^{i(kn-\omega_k \tau)}\hat{a}_k+e^{i(\omega_k \tau -kn)}\hat{a}^{\dag}_k].   
\label{jontau}
\end{equation}
Here, $\hat{a}_k$ and $\hat{a}_{k}^{\dag}$ are Bose operators.
Using these expressions, we calculate the current-current correlator

\begin{equation} 
\Pi (k,\omega)=-i \int_0^\infty \, d \tau  {\rm e}^{i\omega \tau} <[\hat{\j}^{\dag}_k (\tau),
\hat{\j}_k (0) ]>  
\label{curcur}
\end{equation}
and the uniform conductivity
\begin{equation}
\sigma (\omega)=-\frac{1}{\omega} \lim_{k \to 0} {\rm Im} 
\Pi (k,\omega)=2\pi e^2t\delta (\omega)
\label{sigma}
\end{equation}
Furthermore, the K/AK pair correlator exhibits quasi-long range order,
 $< \hat{A}^{+}_{n}\hat{A}^{-}_{m}> =
<\exp  i ( \hat{\varphi}_n - \hat{\varphi}_m )> \propto |n - m|^{- \gamma}$, 
with $\gamma = \sqrt{J/8 \pi^2 t}$. 
Hence, in the limit $t \gg J$ the average dimension of the K/AK pairs diverges (the pairs
decouple), providing the conducting GS.  However,  this does not mean the absence of the 
correlations. Controversally, the GS exhibits quasi-long range correlations of the phase.

The most important property of the GS in this limit  is the presence of decoupled kinks
and antikinks, which form a new kind of correlated state: a {\it kink-condensate}.
Another property of the kink-condensate is its gauge invariancy. Hamiltonian
(\ref{seg}) possesses the gauge symmetry $(\varphi_n \rightarrow \varphi_n + a)$, which is
generated by the topological charge operator $\hat{Q}=-i\sum_n \partial /\partial \varphi _n$ .
Since the kink-condensate is the single GS, it is gauge-invariant.  This can be also seen from
the fact that the total  topological charge is $Q=0$.  The phonon-like excitations (\ref{phonon}) 
break the gauge symmetry and do not break the translational one $(x_n \rightarrow x_n + a)$.
They obey the  Landau criterion \cite{Land}, providing superflow and supercurrent for 
sufficiently small velocities of the condensate $v < v_0 =(2/\pi) \sqrt{2tJ}$.  In this way,
the superconductivity of the quantum string is related to the existence of the kink-condensate.

It is important to notice that in the case of the Josephson chain, the variable $\varphi_n$
is a single-valued function of $n$. In order to keep it single-valued after the dual
transformation (\ref{dt}), one has to require the total momentum of the string to be zero,
$P = \sum_n p_n = 0$. Otherwise, the periodic condition $\varphi_{N+1} = \varphi_1$ cannot
be satisfied and the phase becomes multivalued. Formally, the restriction $P=0$ excludes
uniform current states of the string. However, the thermodynamical limit $N \to \infty$
allows to treat the uniform conductivity  as a long-wave limit  $k \to 0$, see 
Eq.(\ref{sigma}).  On the other hand, the uniform current state describing motion of
the condensate with velocity $v$ can be generated by applying the operator 
$\exp[i(v/2t)\sum_n \hat{x}_n]$ on the GS of the condensate \cite{Froh}.

Finally, we can summarize our results: at $t = 0$, the GS of the string is the
kink-vacuum. At $0<t/J \ll 1$, it has only bound K/AK pairs, exhibits no long-range phase
order, and the energy spectrum is gapped. Then the system is insulating.  At $t/J \gg 1$, the
K/AK pairs are already decoupled and there is no gap anymore. Then the phase is quasi
long-range ordered, the GS is the kink-condensate and the system is superconducting. 
Thus, our
calculations in both limitting cases are in agreement with the results obtained from the
mapping onto the Josephson chain, with the advantage that they clarify the physical meaning of
the insulating and superconducting states for the quantum string. 

Based on the Josephson chain results, it follows that at $(t/J)_c = 2 / \pi^2$
the quantum string undergoes a KT-transition. This transition has
been qualitatively predicted by Eskes {\it et al.} \cite{Eske}, and treated as roughenning 
of the string. Besides, Vierti{\"o} and Rice \cite{Vier} have calculated the energy for creating 
a pair K/AK and have shown that for large $t/J$ values this energy becomes negative, leading to a
proliferation of K/AK pairs. 
Here, we have shown that at the transition point the gap 
$\Delta$ vanishes and the K/AK pairs decouple. However, this decoupling means not only
roughenning of the string, but also {\it quantum depinning} from the lattice and 
{\it superconductivity} of the rough depinned phase. The quantum fluctuations turn out
to be able to depin the string from the lattice and to restore the  Fr{\"o}hlich's perfect
conductivity. Although our conclusions are based on the single stripe picture, we expect the
results to remain valid at higher (but not too high) doping concentrations, as well as in the
presence of impurity pinning. Indeed, as shown in Ref.\ \cite{Hass}, by accounting for both
factors, a ``free phase'' arises within a certain parameter range.  Actually, we show that this
``free phase'' is superconducting.  

We want to emphasize that at 
finite temperatures $(T \ne 0)$, thermal fluctuations will ``spoil'' the
superconductivity for both, the Josephson chain and the quantum string. 
In this case, the Euclidean action (\ref{Ecaction}) describes a XY model on
a 2D lattice, which is finite in the $\tau$-direction, with length 
 $L = 2 \pi /T$. Then, the KT-transition disappears and the long-range
phase correlations, as well as superconductivity, are suppressed. This problem is 
rooted in the 1D treatment of the dynamics. Since we are considering a filled 
stripe and accounting only for transversal fluctuations, superconductivity 
cannot take place, except at $T = 0$. This could explain why the nickelates
never become superconducting. At higher dimensions, thermal fluctuations do
not play such a destructive role anymore. 
Another problem is the Meissner effect and flux quantization. They cannot be verified in 
the 1D model, due to the absence of non-zero volume and of small contours, respectively.
In 1D there is no difference between superconductivity and perfect conductivity. 
Therefore, a 2D theory, coupling the
longitudinal and transversal dynamics in a half filled stripe seems to be
the next step needed for gaining some insight about superconductivity in the
cuprates.

We are indebt with N.\ Hasselmann, H.\ Schmidt, T.\ M.\ Rice and A.\ H.\ 
Castro Neto for fruitfull
discussions. Yu.\ A.\ D.\ acknowledges financial support from the Otto Benecke-Stiftung. 
This work was also partially supported by the DAAD-CAPES project number 415-probral/sch{\"u}.

\end{document}